\documentclass[twocolumn,aps,showpacs,prb,epsf,graphics,psfig]{revtex4}
\usepackage{graphicx}
\usepackage{graphicx}
\usepackage{bbold}
\usepackage{color}
\usepackage[normalem]{ulem}
\usepackage{amssymb}

\begin{document}
\title{
A DNA damage multi-scale model for NTCP in proton and hadron therapy
}

\author{Ramin Abolfath$^{\dagger}$, Chris Peeler, Dragan Mirkovic, Radhe Mohan, David Grosshans$^{*}$}
\affiliation{
Department of Radiation Physics and Oncology, University of Texas MD Anderson Cancer Center, Houston, TX, 75031, USA 
}

\date{\today}

\begin{abstract}
{\bf Purpose}: To develop a first principle and multi-scale model for normal tissue complication probability (NTCP) as a function of dose and LET for proton and in general for particle therapy with a goal of incorporating nano-scale radio-chemical to macro-scale cell biological pathways, spanning from initial DNA damage to tissue late effects.

{\bf Methods}: The method is combination of analytical and multi-scale computational steps including
(1) derivation of functional dependencies of NTCP on DNA driven cell lethality in nanometer and mapping to dose and LET in millimeter, and (2) 3D-surface fitting to Monte Carlo data set generated based on post radiation image change and gathered for a cohort of 14 pediatric patients treated by scanning beam of protons for ependymoma. We categorize voxel-based dose and LET associated with development of necrosis in NTCP.

{\bf Result}: Our model fits well the clinical data, generated for post radiation tissue toxicity and necrosis. The fitting procedure results in extraction of in-{\it vivo} radio-biological $\alpha$-$\beta$ indices and their numerical values.

{\bf Discussion and conclusion}: The NTCP model, explored in this work, allows to correlate the tissue toxicities to DNA initial damage, cell lethality and the properties and qualities of radiation, dose and LET.
\end{abstract}
\maketitle
\section{Introduction}
It is highly challenging in radiation therapy to treat patients and target tumors without involving and risking normal tissues in the surrounding volumes.
Optimization of patient treatment plans allows minimizing potential harms to normal structures and in particular critical tissues while maximizing the conformity of prescribed radiation dose.

Hence, optimization processes of the patient treatment plans rely on two objectives: (1) maximizing tumor control probability (TCP) and (2) minimizing normal tissue complication probability (NTCP).

In proton therapy, and in general, particle therapy, deposited dose and LET are two mechanistic variables that TCP and NTCP depend on.
Any (personalized) model as such must constitute radio-biological responses of tissues throughout hierarchical stochastic and complex pathways that starts from initial DNA damage and subsequently being expressed in cell death machinery~[\onlinecite{Borges2008:CR}].

\begin{figure}
\begin{center}
\includegraphics[width=1.0\linewidth]{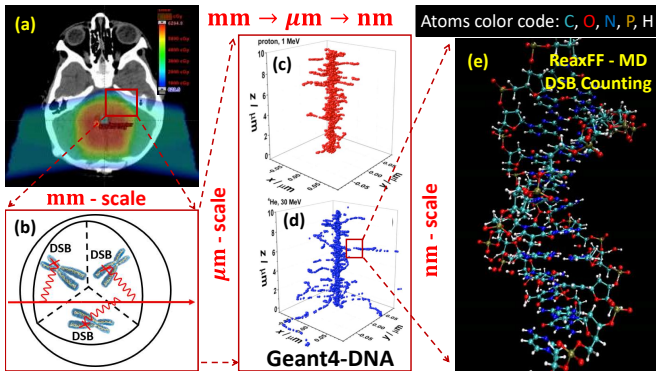} 
\noindent
\caption{
Schematic diagram of multi-scale processes from millimeter to micro- and nanometer.
Shown in (a) a typical patient DICOM RT image in cm-scale.
(b) Passage of a charged particle (red bold arrow) through a cell nucleus in $\mu$m-scale.
Shown in (c) and (d)
the track structures of the ionization process in nano-meter scale.
Shown in (e) a transient structure of damaged DNA surrounded by water molecules and oxygens.
For clarity in visualization, the water and surrounding oxygen molecules are removed from the computational box.
Atoms color code: carbon, oxygen, nitrogen, phosphorus and hydrogen atoms
are shown as cyan, red, blue, gold and white, respectively.
}
\label{fig1_multi_scale}
\end{center}\vspace{-0.5cm}
\end{figure}

Thus a multi-objective optimization technique rely on precise knowledge in spatial distribution of dose and LET in tumors and surrounding tissues, implemented in a reliable and accurate multi-scale radio-biological response theory and model.
The gold standard in retrieving this information in former is the Monte Carlo simulation of clinical beams over patient's data and CT images
~[\onlinecite{Titt2008:PMB,Sawakuchi2010:MP}].
The latter (radio-biological response theory) requires developing multi-scale models that span over several orders of magnitude in time and space, i.e., from $10^{-18}$s to $10^8$s and 1 nm to 1 cm respectively, to capture the post radiation physiological pathways from initial DNA damage to tissue late effects as the current NTCP models ~[\onlinecite{Niemierko1991:RO,Kallman1992:RB,Zaider1999:IJROBP,Thames2004:IJROBP,Peeler2016:RO}] do not capture adequately these essential biological pathways in great details, in particular the dependencies on LET.
A model as such, was recently developed and used for the present analysis as shown in Fig.~\ref{fig1_multi_scale}.

It is known that the relative LET is typically high in low dose volumes in tissues such as penumbras in open fields and regions proximal and distal to Bragg peaks.
However, whether or not variabilities in LET is one of the predictors of NTCP (in addition to dose) is still under scrutiny. 
Besides practical challenges in low dose measurement techniques and scoring post-radiation phenotypes in medical images and correlating these effects to in-{\it vivo} radio-biological pathways, the current NTCP models
[\onlinecite{Niemierko1991:RO,Kallman1992:RB,Zaider1999:IJROBP,Thames2004:IJROBP,Peeler2016:RO}]
lack adequate details in offering conclusive and precise connections to LET.
The present study is an attempt to fill the gap between NTCP and LET from a first-principle point of view.

In this study, we report developing a multi-scale model for NTCP.
We revisit recent publication from our group [\onlinecite{Peeler2016:RO}] that reported selective analysis of MRI-T$_2$ FLAIR images of pediatric patients treated for ependymoma.
Analysis of the images by Peeler {\em et al.} [\onlinecite{Peeler2016:RO}] aimed to correlate the post-radiation necrosis in domains in normal tissues, beyond the tumor location, i.e., in brainstem, and to unveil the effect of unwillingly exposed low doses and high LETs and the proton RBE spatial variabilities to tissue toxicities.
The present NTCP model is a refined and improved version the model employed by Peeler {\em et al.} [\onlinecite{Peeler2016:RO}], as it considers tissue and cell death through DNA damage-activated death machineries that initiates apoptosis and necrosis pathways [\onlinecite{Borges2008:CR}].
We point out that necrotic or radiation induced CNS toxicity may not depend entirely on DNA damage as neurons are post mitotic, however DNA damage likely serves as a surrogate for additional as yet unknown, cellular toxicities.

Our model calculation starts from recently developed multi-scale computational platform that calculates a probability in tissue toxicity, $P$, as a function of dose and LET for scanning beam of protons~[\onlinecite{Abolfath2017:SR,Abolfath2019:EPJD}].
In this model we derived an analytical expression for $P$ based on a system of Markov chains in vastly different spatio-temporal landscapes.
The model starts from the time evolution of DNA lethal lesions that is the driving force, and an input to $P$.
Computationally, we simulate the initial DNA damage in nano-meter (nm) from event-by-event energy loss processes that leads to chromosome aberration and subsequently transfer the results to the cell deactivation in micro-meter ($\mu$m) and tumor response in millimeter (mm).
Scaling up the processes from nm to mm in space, and $10^{-18}$s to $10^8$s in time, we derive an explicit dependence of $P$ on dose and LET in a model that constituents essential radio-biological pathways.

In this study, we start with a general expression for $P$ and calculate NTCP. We then fit P and NTCP to both in-{\it vitro} and in-{\it vivo} dataset. We then show that fitting to clinical data allows an extraction of numerical values of in-{\em vivo} $\alpha$ and $\beta$ radiobiological indices.

This paper is organized as following: We present the analytical steps of the model calculation in Sections \ref{Sec_Model_Outline} to
\ref{Sec_P}, in addition to the analytical relation between NTCP and DNA-DSBs that will be discussed in Sec. \ref{Sec_NTCP_SF}.
In Sections \ref{Sec_Results_invitro} and \ref{Sec_Results_invivo} 
we illustrate the numerical results based on cell survival assay in-{\it vitro} and the 3D-surface fitting of the clinical data, respectively.
We conclude our work in Sec. \ref{Sec_Conclusion}.


\section{Materials and Methods}
\subsection{Outline of model calculation}
\label{Sec_Model_Outline}
\begin{figure}
\begin{center}
\includegraphics[width=1.0\linewidth]{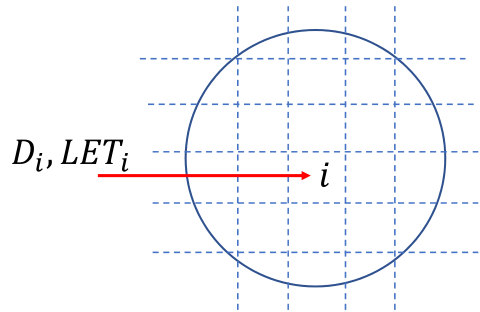}\\ 
\noindent
\caption{
Schematically shown a tissue divided into $N$ segments or voxels.
In each segment, a deposited dose, $D$, and LET is scored.
}
\label{fig1}
\end{center}\vspace{-0.5cm}
\end{figure}

Fig.~\ref{fig1} schematically shows segments of a normal tissue, partitioned into $N$ virtual domains.
For brevity we call these domains, voxels.
The level of radiation exposure and direction of the beam to these normal tissue voxels vary randomly.

As particles deposit energy, a sequence of random damages in tissue may occur.
In this work we consider class of cell damages through apoptotic / necrotic pathways, initiated by formation of DNA damage including double strand breaks (DSB) and more complex cross-links [\onlinecite{Borges2008:CR}], that in turn trigger chromosome abberations.
%
%

To calculate spatially averaged dose and LET, denoted by $D_i$ and $LET_i$, respectively, and simulate our clinical proton beams in each voxel, we used our in-house Monte Carlo (MC) toolbox.
Here $i$ is an integer number that allows enumeration of the voxels.

We assign a probability $P_i$ to tissue injury event in voxel $i$th,
with a pair of labels, 0 and 1, to denote occurrence of post-radiation image change.
As we consider a binary sample space in scoring random processes in tissue injuries, $1 - P_i$ yields a probability in no-injury.
To clarify nomenclature and terminology in this study, we consider tissue injury in a voxel if post-radiation image change in the same voxel is observed.
Note that the occurrence in tissue injury depends on the severities in DNA and cell damage, hence $P_i$ depends on $D_i$ and $LET_i$.
We further consider $P_i$ a continuous function of dose and LET such that in absence of radiation, e.g., $D_i = LET_i =0$, no radiation injury takes place, hence $P_i = 0$.
Moreover, $P_i$ increases monotonically from zero to unity as $D_i$ and $LET_i$ increase.
The details in steps and our formulation in calculating $P_i$ is given in the following sections.

\subsection{Monte Carlo}
\label{Sec_Model_MC}
As described by Peeler {\em at al.} [\onlinecite{Peeler2016:RO}], Monte Carlo (MC) techniques were used to produce dose and LET distributions for the patient treatment plans. The plans used for the study were originally generated in the Eclipse treatment planning system (v9.0) and were subsequently recalculated using an MCNPX-based in-house MC system in order to produce dose and fluence distributions. The MC system takes as input the DICOM files from the TPS which it uses to create geometry specifications as well as files defining the beam arrangement and delivery. The design of the system has been described previously [\onlinecite{Titt2008:PMB,Sawakuchi2010:MP}]. These quantities were then used to calculate track-averaged LET distributions for each of the cases. For any given plan, which may have consisted of a primary and one or more boost plans, the dose was calculated as a sum and the track-averaged LET was calculated as an average over all plans.

Track-averaged LET was utilized in MCNPX-based MC calculation because it would have been prohibitively time consuming to produce the data required to calculate dose-averaged LET in each voxel, as this would have required the scoring of the energy fluence spectrum in every voxel. The LET values, calculated from LET distributions and typically observed in full treatment plan are low. In this region the track-averaged LET can be an acceptable approximation of the dose-averaged LET. For higher values of LET, the dose-averaged LET is generally greater than the track-averaged LET~[\onlinecite{Guan2015:MP}]. Were dose-averaged LET to be used for the analysis, we would thus expect to observe somewhat higher values of LET for those values already on the higher end of the spectrum.

\subsection{Constant $P$ over ensemble of voxels}
\label{Sec_Model_Const_P}
If all voxels were identical, we could remove the index $i$ from $P_i$ and consider a constant probability $P$ for tissue injury over the entire tissue volume.
Although this is not a realistic assumption for a human tissue under radiation of a clinical beam, but we may consider it as a hypothetical limiting case that simplifies calculation of NTCP.

Assertion of this assumption turns the actual tissue response to binary-events of exposing $N$ copies of identical voxels to a given dose and LET, similar to flipping copies of $N$ identical coins (with probabilities $P$ and $1-P$ in scoring head and tail) as $P_i$ is independent of voxel index $i$.
This is regardless to a location of a voxel with respect to the beam source, spot size geometry, the details in energy-loss and lineal energy distributions, in addition to inhomogeneities in tissue physiological structure.
Thus this assumption might be reasonable for NTCP in outer space where there is no consideration in radiation source position with respect to location of tissue.

With further assumption of weakly correlated tissue structures in adjacent voxels, the probability in finding a tissue with injury can be calculated by a binomial / Bernoulli distribution
\begin{eqnarray}
N_n = \left(\begin{array}{cc} N \\ n \end{array}\right) P^n (1 - P)^{N - n}.
\label{eq1}
\end{eqnarray}
$N$ is the total number of segments / voxels of the normal tissue and $n$ is the number of voxels scored with occurrence of injury. $N_n$ is a normalized distribution function, hence $\sum_{n=0}^{N} N_n =1$.

Let us consider series expansion of the normalization condition and recall the following well-known identity of the binomial distribution
\begin{eqnarray}
1 &&= \sum_{n=0}^{N}N_n =
\sum_{n=0}^{N} \left(\begin{array}{cc} N \\ n \end{array}\right) P^n (1 - P)^{N - n}\nonumber \\&&
= (1 - P)^N + N P (1- P)^{N - 1} \nonumber \\&&
+ \frac{N(N-1)}{2!} P^2 (1 -P)^{N-2}
+ ... + P^N.
\label{eq2}
\end{eqnarray}
The first term in Eq.(\ref{eq2}) represents the probability that all $N$-elements of the tissue escape from injury.
Similarly the last term represents the probability that all $N$-elements of the tissue experience injury.
The rest of terms in Eq.(\ref{eq2}) are between these two extreme limits.
For example, the second and third terms in Eq.(\ref{eq2}) represent the probabilities that exactly one and two elements experience injury and the rest of elements experience no injury.

In this work, we define NTCP as the probability in which at least one element of the normal tissue experiences injury.
This corresponds to all terms in Eq.(\ref{eq2}), except the first one, hence
\begin{eqnarray}
NTCP = 1 - (1 - P)^N.
\label{eq2}
\end{eqnarray}
Eq.(\ref{eq2}) is consistent with the NTCP model introduced by Niemierko and Goitein in Ref.~\onlinecite{Niemierko1991:RO}.
One may extend this approach and construct a cluster / percolation model for NTCP with a minimum number of injured-elements (see for example, Ref.~\onlinecite{Thames2004:IJROBP}).
We do not consider this extension in the present work.

In a careful treatment plan, optimized well to lower the probability in normal tissue injury, $P$ is a small number hence
\begin{eqnarray}
NTCP = \sum_{n=1}^N N_n = 1 - (1 - P)^N \approx N P.
\label{eq22_0}
\end{eqnarray}
We refer to RHS of Eq.(\ref{eq22_0}), a linearized model of NTCP.

\subsection{Graded-$P$}
\label{Sec_Model_Graded_P}
In a more clinically relevant model of NTCP, $P$ is a continuous function of dose and LET. Therefore one may apply the transformation
$(1 - P)^N \rightarrow \prod_{i=1}^{N} (1 - P_i)$ in Eq.(\ref{eq22_0})
and extend the NTCP to the following formulation
\begin{eqnarray}
NTCP = 1 - \prod_{i=1}^{N} (1 - P_i(D, LET)).
\label{eq22}
\end{eqnarray}
In a situation where in a given voxel, e.g., $jth$ voxel, $D = LET = 0$, $P_j = 0$, hence
$NTCP = 1 - \prod_{i\neq j}^{N} (1 - P_i(D, LET))$.
Therefore if $N - N_x$ voxels experience no dose and LET we can express
\begin{eqnarray}
NTCP = 1 - \prod_{i=1}^{N_x} (1 - P_i(D, LET)),
\label{eq23}
\end{eqnarray}
where $i$ runs over subset of voxels that experience receiving any amount of dose and LET.
Here $N_x$ is the number of voxels, accounted for deposition of dose and LET.
Note that in voxels receiving small amount of doses including in penumbras of open fields and volumes proximal / distal to Bragg peaks, where LET is expected to be high, $P(D, LET)$ should be large if $LET$ is one of the predictors of NTCP.

To the best of our knowledge, in NTCP models available in literature, see for example Refs. [\onlinecite{Niemierko1991:RO,Kallman1992:RB,Zaider1999:IJROBP,Thames2004:IJROBP,Peeler2016:RO}],
no explicit dependence of NTCP on LET were reported.
Thus a systematic derivation of $P(D, LET)$ is needed to be performed.
The following section is devoted to sketch careful derivation of $P$ as a function of dose and LET.

\section{Calculation of $P$}
\label{Sec_P}
Center to this model is the calculation of $P$,
a probability that a voxel in normal tissue experiences post-radiation image change as a result of the cell lethal injury in that voxel.
We note that our approach in calculating $P$ is different from the approach introduced by Niemierko and Goitein~[\onlinecite{Niemierko1991:RO}] and the subsequent publications intended to refine their original approach in photon therapy (see for example Ref. \onlinecite{Zaider1999:IJROBP}).

We define $P$, the probability in which all of the cells in a normal-tissue element, $N_0$, experience lethal injury.
Note that, we have no rigorous evidence, neither experimentally nor theoretically, to verify if lethal injury of all cells in a voxel is a necessary condition for the occurrence of the post-radiation image change in that voxel.
It is possible that occurrence of cell lethality in a subset of $N_0$ cells, e.g., number of cells less than $N_0$, would be a sufficient condition for scoring post-radiation image change.
However, because of simplicity in computation, we assume all cells in a voxel participate collectively in post-radiation image change.
Alternatively, one may consider $N_0$ an additional fitting parameter that varies within 0 and the maximum number of cells in a voxel.
Such variation can be implemented voxel by voxel, however, it introduces substantial complication in computation of NTCP, hence we do not consider
the effect of variability in cell lethality per voxel and its correlations to post-radiation image change in our current study. We postpone to investigate it in future.

To proceed with analytical derivation of $P$, we first divide the normal tissue element into millimeter, then to micrometer and finally to nanometer size virtual domains as shown schematically in Fig.~\ref{fig1_multi_scale},
a representation of our multi-scale computational platform.
In Fig.~\ref{fig1_multi_scale}(a) we show a typical patient DICOM RT image in cm-scale.
The patient was treated for ependymoma by a pair of lateral scanning proton beams technique at the proton center in our institution. The deposited dose depicted in color codes from low (blue) to high (red) doses.
In Fig.~\ref{fig1_multi_scale}(b), passage of a charged particle (red bold arrow) through a cell nucleus in $\mu$m-scale is shown. DNA-materials and chromosomes in cell nucleus undergo complex damages such as DSBs.
The wiggly lines are photon field propagators, a Feynman diagram representation in quantum electrodynamics (QED)
that describes interaction of charged particles in scattering processes with a random site.
The site of interaction is either on DNA and/or the environment of DNA in cell nucleus.
It describes disintegration process of water molecules into Hydrogen and release of OH-free radicals and/or oxygen molecules into reactive oxygen species (ROS) in indirect DNA damage (the latter), or release of
shell electrons localized initially in DNA in direct damage processes (the former).
In Fig.~\ref{fig1_multi_scale} (c) and (d)
the track structures of the ionization process in nano-meter scale are shown. The ionization tracks were calculated by using Geant4-DNA Monte Carlo toolkit~[\onlinecite{Incerti2010:IJMSSC}], for 1 MeV proton and 30 MeV He respectively in water.

Fig.~\ref{fig1_multi_scale} (e) shows a transient structure of DNA surrounded by water molecules and oxygens  at room temperature after 10 ps ReaxFF molecular dynamics~[\onlinecite{Abolfath2011:JPC,Abolfath2013:PMB}]
simulated in fraction of femto-second time steps right after ionization process shown in (b), (c) and (d).
For clarity in visualization, the water and surrounding oxygen molecules are removed from the computational box.
In each segment a fragment of DNA is fitted. The DNA fragment is long enough to simulate DNA damage including DSB lesions.
A typical snap-shot of distorted DNA, as seen in Fig.~\ref{fig1_multi_scale} (e), contains one or more single strand breaks (SSB).
SSBs located within 10 base-pairs separation are considered formation of a DSB. In this figure, damaged bases,
distorted hydrogen bonds, base-stacking are seen.
Although not shown in this figure, such atomistic simulations reveal formation of complex ROS in cell nucleus (in the environment of DNA) such as hydrogen-peroxides, formed from two OH free radicals in addition to more complex oxygen-hydrogen compounds.
Atoms color code: carbon, oxygen, nitrogen, phosphorus and hydrogen atoms
are shown as cyan, red, blue, gold and white, respectively.

Thus the multi-scale steps in calculating number and types of damages in DNA consists of first performing a collision by collision following by event by event Geant4-DNA MC simulation~[\onlinecite{Incerti2010:IJMSSC}] of track structures in a voxel in normal tissue volume, as illustrated in Fig.~\ref{fig1_multi_scale}(b)-(d).
In second step, we perform a molecular dynamic simulation for scoring initial structural damages to DNA induced by ionization of surrounding water and oxygen molecules to OH$^\cdot$ free radicals and reactive oxygen species~[\onlinecite{Abolfath2011:JPC,Abolfath2013:PMB}]. See Fig.~\ref{fig1_multi_scale}(e).
Subsequently, the nm-events, including DNA-DSBs are averaged out to scale up the lethal lesions at the cellular level to simulate the cell survival fraction. See Fig.~\ref{fig1_multi_scale}(b).

The steps involving these algebra is cumbersome but it is straightforward to show~[\onlinecite{Abolfath2019:EPJD}]
\begin{eqnarray}
P\left[\overline{n}(t)\right] = \left(1 - SF_{\rm eff} \left[\overline{n}(t)\right]\right)^{N_0}.
\label{eq002cds3}
\end{eqnarray}
Here $\overline{n}(t)$ is the average number of DNA DSB's in a cell nucleus in a voxel (for simplicity in notations we skip carrying voxel index $i$), a variable that is function of time, $t$. $N_0$ is number of cells that constitute a normal tissue voxel, and
\begin{eqnarray}
SF_{\rm eff}\left[\overline{n}(t)\right]
= \frac{e^{(b - d)t-\overline{L}\left[\overline{n}(t)\right]}}{1 + b e^{(b - d)t-\overline{L}\left[\overline{n}(t)\right]}
\int_{0}^{t} dt' e^{-(b - d)t' + \overline{L}\left[\overline{n}(t')\right]}},
\label{eq002cds3f}
\end{eqnarray}
is effective cell survival fraction (SF). In Eq.~\ref{eq002cds3f}, $b$ and $d$ denote the birth and death rates of cells in absence of radiation field and
\begin{eqnarray}
\overline{L}\left[\overline{n}(t)\right] = \int_0^t dt' \left(\lambda_{L,{\rm eff}} \overline{n}(t') + \gamma_{L,{\rm eff}} \overline{n}^2(t') + ...\right),
\label{eq0027zz}
\end{eqnarray}
is average number of lethal lesions in a cell nucleus.
$\overline{L}$ is a functional integral of $\overline{n}(t)$, the average number of DSBs in a cell that resembles the driving force in $P$, i.e., the higher $\overline{n}$, the closer $P$ to one.
$\lambda_{L,{\rm eff}}$ and $\gamma_{L,{\rm eff}}$ are enzymatic misrepair rates corresponding to DNA and chromosomes lethal lesions. They are phenomenological coefficients that can be determined by the fitting to the experimental cell survival data.
As discussed in Ref.~[\onlinecite{Abolfath2019:EPJD}], $\overline{L}$ fits a linear-quadratic type-relation in dose with $\alpha$ and $\beta$, polynomial functions of stochastic lineal-energy, $y$, or its deterministic counterpart, LET, i.e., $\overline{L} = \alpha(LET) D + \beta(LET) D^2 + {\cal O}(D^3)$.
Here $\alpha = \sum_{k=0}^q \alpha_k LET^k$ and $\beta = \sum_{k=0}^{q-1} \beta_k LET^k$ are polynomials with rank $q$ and $q-1$.

These equations exhibit interplay of sequence of events starting from DSB induction by ionizing radiation in microscopic scale that propagate to formation of DNA damage and chromosome misrepairs to a coarse-grained tumor dynamical responses in macroscopic scale.
These processes govern underlying mechanisms that constitute multi-scale formulation of the cell deactivation in proton and hadron therapy.
For more details in mathematical derivations and steps refer to Ref.~[\onlinecite{Abolfath2019:EPJD}].

\subsection{NTCP and SF}
\label{Sec_NTCP_SF}
To simplify the mathematical expressions for calculation of NTCP, we consider a limit where
$SF_{\rm eff}$ can be replaced by $SF$ (e.g., the limit where $b=d=0$, e.g., mature neuron cells), hence
\begin{eqnarray}
P(D, LET) = \left(1 - e^{- \overline{L}\left[\overline{n}(\infty)\right]}\right)^{N_0}
= \left(1 - SF\right)^{N_0}.
\label{eq002cds3xx}
\end{eqnarray}
Considering $N_x$ voxels for normal tissue and recalling Eq.(\ref{eq23})
\begin{eqnarray}
NTCP  &=& 1 - \prod_{i=1}^{N_x}(1 - P_i) \nonumber \\ &&
= 1 - \prod_{i=1}^{N_x}\left(1 - \left(1 - SF_i\right)^{N_0}\right).
\label{eq32}
\end{eqnarray}
we find
\begin{eqnarray}
NTCP &\approx&
1 - \prod_{i=1}^{N_x}(1 - e^{- N_0 \times SF_i}) \nonumber \\ &&
\approx 1 - \prod_{i=1}^{N_x} e^{-e^{- N_0 \times SF_i}}    \nonumber \\ &&
\approx 1 - e^{- \sum_{i=1}^{N_x} e^{- N_0 \times SF_i}},
\label{eq33}
\end{eqnarray}
and finally
\begin{eqnarray}
NTCP \approx
1 - e^{- \sum_{i=1}^{N_x} e^{- N_0 \times e^{-\alpha(LET_i) D_i - \beta(LET_i) D^2_i}}}.
\label{eq34}
\end{eqnarray}

\subsection{Surface fitting}
The evaluation of post-radiation image change and categorizing the voxels were assessed by our internal physician and radiologist.
A binary assignment is given to attach 0 and 1 labels to the voxels.
Voxels with no image change are mapped to zero, and one otherwise.
To calculate dose and LET in each voxel, the patients' DICOM images were plugged in the Monte Carlo code with a series of beam arrangements that resembles the clinical setup in the patient's treatment.
The tissue toxicity is therefore mapped to a two-dimensional space of dose and LET.
We then calculated a histogram for population of voxels that were assigned for tissue toxicity.
This transformed voxel population within an interval of dose and LET to a probability density per voxel, $P_{\rm exp}$, a discrete function of dose and LET.
The histogram, hence $P_{\rm exp}$, was represented in a scatter-plot in a three-dimensional space with coordinates spanned by calculated dose and LET.

In the next step, we fitted a two-dimensional continuous surface, $P(D, LET)$, to the scattered-points, $P_{\rm exp}$, a discrete function of dose and LET.
The fitting procedure was performed under a set of mathematical constraints.
We enforced the fitted surface to follow linear-quadratic model in dose.
In addition, we asserted a set of conditions on the linear-quadratic $\alpha$-$\beta$ indices because they follow a coupled system of power-law series in LET.
Note that, from nano-scale DNA damage and repair model, we found a relation between the ranks of polynomials in $\alpha$ and $\beta$.
Accordingly, $\alpha$ and $\beta$ are polynomials of LET with ranks $q$ and $q-1$ respectively.
We therefore asserted the latter as another constraint in fitting a surface to scatter-plot of tissue toxicity points.
Subsequent to calculation of $P$, we use analytical equations derived in our model and calculate the corresponding NTCP.

The numerical values of the fitting parameters, were calculated within an in house developed Matlab code designed for an optimization algorithm in searching the best 2D surface fitted to a data set in 3D space. An iterative procedure was employed to minimize the chi-square value to obtain the optimal parameter values and performing nonlinear surface fitting. The optimization core of our approach uses the implementation of Levenberg-Marquardt-Fletcher algorithm developed for nonlinear least squares fitting problems.

We consider the polynomial ranks of $\alpha$ and $\beta$, $q$ and $q-1$ as free parameters and fit surfaces as increasing function of $q$ until a satisfactory convergence is obtained.
The rationale for selecting $q\geq 2$ stems from numerical methods in optimization and convergence of the self-consistent $P$-solutions in surface fitting procedure.
By increasing the powers in the expansion, the numerical values of the coefficients will change, however, if the change in $P$ resides within a convergence domain (chosen by the user) the optimization stops.  Although in our approach we search for optimal surfaces to minimize chi-square values iteratively, it is known that additional metrics are needed to evaluate the goodness of the fit. We therefore calculate R-square, also known as coefficient of determination (COD).
The numerical value of R-square for the final fitted surface was close to 1 as the closer the fit is to the data points, the closer R-square to be to the value of 1. It is also known that a larger value of R-square does not necessarily mean a better fit because of the degrees of freedom that can also affect the values. Thus we calculate the adjusted R-square values to account for the degrees of freedom.

Similarly to determine an optimal polynomial in the fitting process of the high-throughput and high accuracy clonogenic in-{\it vitro} cell-survival data, hence $P$, we consider a series of polynomials corresponding to $q=2,3,4, \cdots$ and perform optimization for each individual polynomial separately. A comparison based on chi-square and SF is performed for polynomials with orders $q$ to $q+1$. If the difference between two chi-square’s, associated with SF’s, in steps $q$ and $q+1$ is lower than a threshold value, e.g., $\left|\chi^2(q+1) - \chi^2(q)\right| < \delta$, the optimization stops and chi-square, $SF$ and $P$ corresponding to step $q+1$ are reported.
In low LET’s this procedure converges to a linear polynomial in $\alpha$ as a function of LET, and a constant $\beta$ with no explicit dependence on LET.

\begin{figure}
\begin{center}
\includegraphics[width=1.0\linewidth]{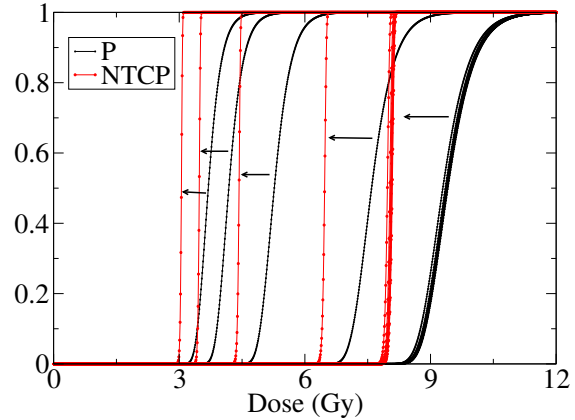}\\ 
\noindent
\caption{
Shown $P$ (black) and the correspondingly $NTCP$ (red) of $10^5$ cells
in a typical mm-size voxel calculated from fitting to an in-{\it vitro} H460 (NSCLC) cell line as reported in Ref.[\onlinecite{Abolfath2019:EPJD}].
$N_x \approx 10^4$ calculated from physical dimensions of the plates and samples used in the experiment~[\onlinecite{Guan2015:SR}].
As LET increases, sigmoid curves shift to lower doses (from right to left).
The arrows indicate the $NTCP$ corresponding to $P$ (same LET curves).
The right most juxtaposed curves correspond to $LET_d = 0.9$ - $5.1 keV/\mu m$. The rest of the curves from right to left correspond to
$LET_d =$ 10.8, 15.2, 17.7 and 19 $keV/\mu m$.
}
\label{fig4}
\end{center}\vspace{-0.5cm}
\end{figure}

\section{Results}
\label{Sec_Results}
\subsection{Illustrative example}
\label{Sec_Results_invitro}
For illustration of our methodology we apply an approximation to Eq.~\ref{eq34} where explicit variation of dose and LET over voxels can be neglected, i.e., the entire tissue irradiated uniformly hence explicit dependence on index $i$
can be discarded
\begin{eqnarray}
NTCP &=&
1 - e^{- \sum_{i=1}^{N_x} P_i(D, LET)} \nonumber \\ &&
\approx 1 - e^{- N_x \times P(D, LET)}.
\label{eq35}
\end{eqnarray}
Here
\begin{eqnarray}
P(D, LET) = e^{- N_0 \times e^{-\alpha(LET) D - \beta(LET) D^2}}.
\label{eq36}
\end{eqnarray}
We note that Eq.~(\ref{eq36}) was suggested to be used in calculation of NTCP by Kallman {\em et al.} [\onlinecite{Kallman1992:RB}]. See for example Eq.~(A5) in Ref.[\onlinecite{Kallman1992:RB}] and the hand waiving arguments used for justifications this equation.
Hence the mathematical details as presented in our study, provides a more rigorous derivation of similar equation introduced by Kallman {\em et al.} [\onlinecite{Kallman1992:RB}].

Fig.~\ref{fig4} shows $P$ and $NTCP$ on the same graph for such a situation.
The numerical values of the coefficients in the polynomial expansion of $\alpha$ and $\beta$ for $LET < 10 keV/\mu m$ turns out to be
$q=1$, $\alpha_0 = 0.21447$, $\alpha_1 = 0.007$ and $\beta_0 = 0.11061$, reduced $\chi^2 = 0.00487$, $R^2$(COD)$=0.99496$ and adjusted $R^2 = 0.99481$.
For $LET > 10 keV/\mu m$, we obtained $q=4$ with
$\alpha_0 = 0.12913$, $\alpha_1 = 6.733 \times 10^{-3}$,
$\alpha_2 = 4.563\times 10^{-4}$, $\alpha_3 = 3.37529\times 10^{-5}$,
$\alpha_4 = 8.16229\times 10^{-6}$ and
$\beta_0 = 0.06864$, $\beta_1 = 0.00202$, $\beta_2 = 1.36897\times 10^{-4}$, $\beta_3 = 3.63748\times 10^{-5}$,
reduced $\chi^2 = 0.007557$, $R^2$(COD)$=0.98407$ and adjusted $R^2 = 0.92833$.

The numerical results shown in Fig.~\ref{fig4}, in particular,
a shift in sigmoid curves to lower doses as LET increases,
mimic the recently experimental data of NTCP and
late normal tissue responses in the rat spinal cord after carbon ion irradiation~[\onlinecite{Saager2018:RO}].
Note that in our NTCP model, the occurrence of the late normal tissue endpoint and the highly structured architecture of normal tissues are a priori known and the complexities in clinical side effects are all embedded in the radiobiological response functions, e.g., in $SF$, $P$ and NTCP.
The simplicity of the present model resides in unification of normal tissues and tumor responses with a unique underlying microscopic pathways regardless of vastly different complexities in normal tissues and tumors responses, i.e., spite of correlations between cell survival and clinical side effects in normal tissues that appears much more complex than for tumors.


\begin{figure}
\begin{center}
\includegraphics[width=1.0\linewidth]{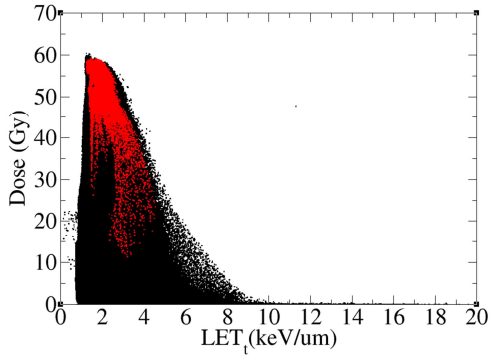}\\ 
\noindent
\caption{
A two dimensional topograph of the clinical data from cohort of 14 patients,
recently presented by Peeler {\em et al.} in Ref.[\onlinecite{Peeler2016:RO}].
$p=0$ (the black dots) identifies the voxels with no post-radiation image change.
Similarly $p=1$ (the red dots) represent the voxels were identified with necrotic post-radiation image change.
}
\label{fig5}
\end{center}\vspace{-0.5cm}
\end{figure}

\begin{figure}
\begin{center}
\includegraphics[width=1.0\linewidth]{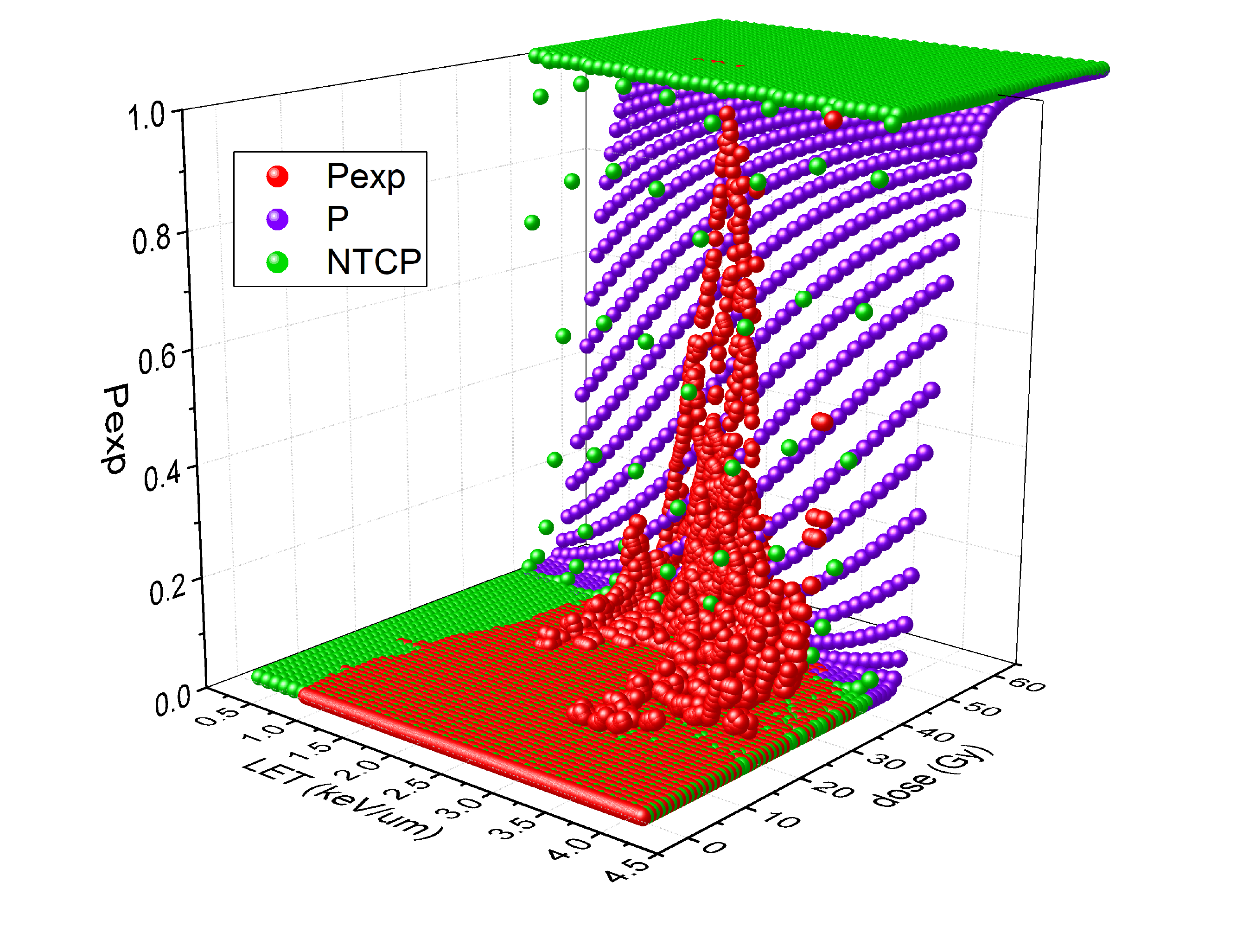}\\ 
\noindent
\caption{
Shown $P$ (blue) and $NTCP$ (green) fitted to the clinical data Peeler {\em et al.} [\onlinecite{Peeler2016:RO}] (red).
$P_{\rm exp}$ is experimental probability density, obtained from histogram of a two dimensional topograph of the clinical data from cohort of 14 patients, presented in Fig.~\ref{fig5}.
}
\label{fig6}
\end{center}\vspace{-0.5cm}
\end{figure}

\subsection{Fitting to ependymoma clinical data}
\label{Sec_Results_invivo}
We now turn to use our NTCP model and present fitting of the clinical data recently reported for evidence of variable proton biological effectiveness in pediatric patients treated for ependymoma~[\onlinecite{Peeler2016:RO}].
Fig. \ref{fig5} shows the voxels based tissue toxicity of cohorts of 14 patients.
The labels 0 (the black dots) and 1 (the red dots) correspond to voxels identified as non-necrotic and necrotic post radiation image change.

The NTCP surface fitted to these clinical data is shown in Fig. \ref{fig6}.
Eqs. \ref{eq35} and \ref{eq36} were used for fitting of the NTCP of the clinical set.
The surface shown in Fig. \ref{fig6} calculated based on the best fitted surface with $N_0 = 8\times 10^6$ cells per voxel where the voxel size is $1.9531 \times 1.9531 \times 2.5 mm$.
We note that an averaged number of exposed voxels for 14 patients was calculated and used for $N_x$.
$\alpha$ and $\beta$ are the output from the fitting of the surface to these data set using
linear LET model, $\alpha = \alpha_0 + \alpha_1 LET$ and $\beta = \beta_0$.
The converged numerical results in the fitting procedure yields
$\alpha_0 = 0.25669$ Gy$^{-1}$, $\alpha_1 = 0.01096$ Gy$^{-1}$/(keV/$\mu$m) and $\beta_0 \approx 0$.

To transform the clinical data from a binary form, presented in Fig.~\ref{fig5}, to a probability, a continuous function of dose and LET, shown by the red dots in Fig.~\ref{fig6}, we counted number of voxels within an interval / bin of dose and LET, $D$ and $D + \Delta D$ and $LET$ and $LET + \Delta LET$ respectively. We then calculate $P = V_1 / (V_1 + V_0)$ where $V_0$ and $V_1$ are number of voxels accounted for no necrosis and with necrosis within these intervals / bins.
$P$ is therefore a normalized histogram calculated for occurrence of necrosis within dose and LET intervals, $\Delta D$ and $\Delta LET$,
the same $P$ as defined in Eq.~\ref{eq36}.
After determining $P$ from fitting procedure (blue dots in Fig. \ref{fig6}) to the clinical data (red dots), we recall Eq.~\ref{eq35} and calculate NTCP (the green dots).

\subsection{Discussion and conclusion}
\label{Sec_Conclusion}
The NTCP model, presented in this work, starts from radiation induced DNA damage in nano-meter and scales up to millimeter tissue damage, observable in medical images.
The methodology introduced in this work allows to apply a general framework to study the radio-biological effects in-{\it vitro} and/or in-{\it vivo}.
We illustrated the applicability of our model and presented the numerical results of NTCP for both cell lines.
The novelty of the present approach relies on correlations among the tissue toxicities, cell lethalities and radiation induced DNA damage, besides dependencies on the radiation dose and LET.

One may critically raise questions on model validation, the computational components how to validate such model, inter-patients statistics and the small cohorts of 14 patients that might not be adequate for the model validation.
It is also important to note that our present analysis is based on mixture of patients voxels, sorted based on the scored dose and LET per voxel. With 14 patients, there are large number of voxels to analyze. This can be seen in Fig. \ref{fig5} where the points are highly dense due to large number of voxels used for fitting the data to a surface.
For a patient by patient model validation, in our preliminary fitting procedure, we divided 14 patients into two pools of training and validation dataset, e.g., 11 and 3, 12 and 2, or 13 and 1, respectively. In the training dataset, we performed NTCP fitting by mixing of the voxels of patients in the training dataset and subsequently tested the result against the mixed voxels of the second pool to validate the approach.
The final result presented in this manuscript, though, is based on fitted NTCP of all 14 patients, simply because we are limited to small number of patients.
Because this work is in progress, in the future, the model will be validated in larger patient cohorts using published techniques and
we may be able to present the results based on two complete separate pools of training and validation.

{\bf Acknowledgement:}
The authors would like to acknowledge useful discussion and scientific exchanges with Dr. Harald Paganetti.
The work at the University of Texas, MD Anderson Cancer Center was supported by the NIH / NCI under Grant No. U19 CA021239.

\noindent{\bf Authors contributions:}
RA: wrote the main manuscript, prepared figures, performed mathematical derivations and computational steps including three dimensional surface fitting to the experimental data.
CP: wrote the main manuscript and contributed to computational analysis including in house Monte Carlo simulations.
DM: contributed to computational analysis including in house Monte Carlo simulations.
DG: provided patient's data and its interpretation and wrote the manuscript and co-supervised the project.
RM: wrote the main manuscript and proposed scientific problem and co-supervised the project.

\noindent{\bf Competing financial interest:}
The authors declare no competing financial interests.

\noindent{\bf Corresponding Authors:}\\
$^\dagger$ ramin1.abolfath@gmail.com \\ 
$^*$ dgrossha@mdanderson.org

\end{document}